\documentclass[amsmath,amssymb]{revtex4}
\usepackage[cp1251]{inputenc}
\usepackage[english]{babel}
\usepackage{color}
\usepackage{graphicx}
\usepackage{braket}
\usepackage{amsmath}
\usepackage{array}
\usepackage[cp1251]{inputenc}
\usepackage{amssymb }
\usepackage{natbib}

\begin{document}

\bibliographystyle{unsrtnat}

\title{Quantum teleportation in terms of creation operators of photons}

\author{M.V. Fedorov{$^{1,\,2\,*}$ and A.A. Sysoeva{$^{1,\,3}$}}}
\address{{\em{$^{1}$A.M.~Prokhorov General Physics Institute, Russian Academy of Sciences, Moscow, Russia}}\\
{\em{$^{2}$ National Research University Higher School of Economics, Moscow, Russia}}\\
{\em{$^{3}$ Moscow Institute of Physics and Technology, Dolgoprudny, Moscow Region, Russia}}\\
$^{*}$fedorovmv@gmail.com}

\date{\today}

\begin{abstract}
By using the formalism of photon creation operators, we present the simplest description of the effect of quantum teleportation and describe its closest classical analog.
\end{abstract}


\maketitle

\section{Introduction}
Quantum Teleportation (QT) is a rather well known and very popular phenomenon described for the first time by Bennett \cite{Bennett} and observed experimentally in various schemes in the works \cite{Zeilinger}, \cite{DeMartini}, \cite{SPK}. Of course, QT has nothing in common with teleportation from fiction books. Instead of transporting objects in space and time, QT consist in providing conditions for the recipient (Bob) to reproduce a copy of the signal obtained originally from somewhere by a sender (Alice). To reach this, at first Alice and Bob receive one photon each from entangled pairs of photons produced in the process of Spontaneous Parametric Down-Conversion (SPDC). Then Alice compares her SPDC photon with that coming in the signal and, based on the results of this comparison, sends instructions to Bob (by phone) on how he has to modify his own SPDC photon to make it identical to the signal one.  Theoretical explanations of QT are based often on the use of photon's wave functions (even if they  are written in the Dirac notations as $\ket{\uparrow_1}$, $\ket{\downarrow_2}$) etc. Such description can be somewhat confusing because the symmetry requirements for the wave functions of three identical particles (bosons or fermions) are not always completely satisfied. On the other hand, if these requirements are completely satisfied, the description becomes rather cumbersome. An alternative way consists in using only quantum-electrodynamical state vectors easily expressed in terms of photon creation operators. As shown below (section 2) such description is so simple that it can be referred to as the ``quantum teleportation for pedestrians". In terms of quantum-electrodynamical creation operators one does not have to bother about symmetry requirements because they are satisfied automatically in the well known rules for constructing multiphoton wave functions from state vectors \cite{Schwe}. In section 3 we will describe a classical analog of QT and discuss the degree of ``quantumness" of QT.

\section{QT in terms of creation operators}

\begin{figure}[h]
\centering\includegraphics[width=12cm]{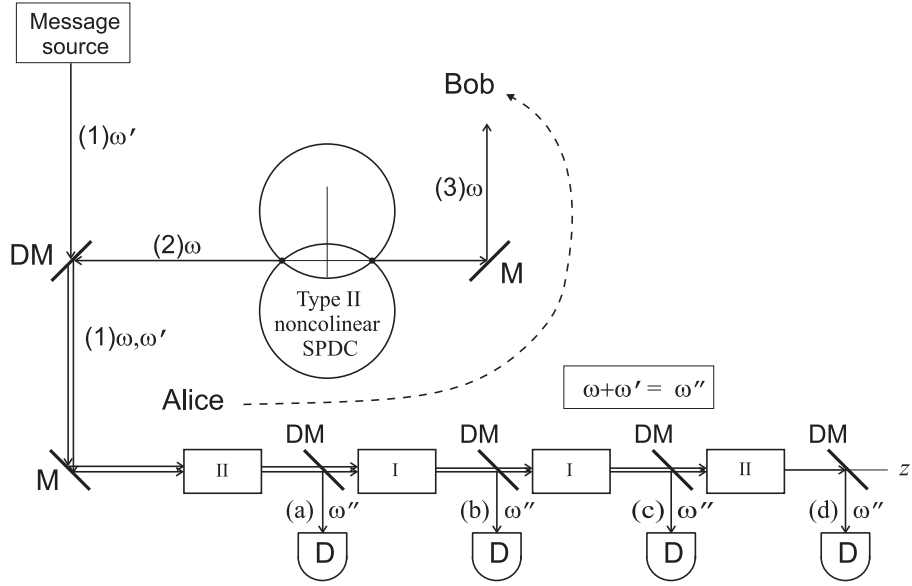}
\caption{{\protect\footnotesize {The QT scheme under consideration; D are detectors, M - mirrors, DM - dichroic mirrors; (1), (2), (3), and ${\rm (a),\,(b),\,(c),\,(d)}$ are numbers and indicators of photon routes, I and II characterize types of phase matching in the frequency-summing crystals}}}\label{Fig.1}
\end{figure}

In this section we present the creation-operator analysis of QT for the scheme of the work \cite{SPK} with slight modifications and simplifications (see  Figure \ref{Fig.1}).
SPDC photon pairs are assumed to be produced in a nonlinear crystal (not shown explicitly) tuned for the noncollinear frequency-degenerate regime with the type-II phase matching. As known \cite{Kwiat}, in this case SPDC photons propagate along two cones corresponding to different polarizations of photons (horizontal, $H$, and vertical, $V$). Two circles in Figure \ref{Fig.1} are sections of the cones by a plane perpendicular to the pump-propagation direction. Only photons from two crossing points of the circles are assumed to be used, with photons from one crossing point sent to Alice and from the other one to Bob. Frequencies of all SPDC photons ($\omega$) are equal to each other and are equal to a half of the pump frequency $\omega=\omega _p/2$. Emitted SPDC photons have two degrees of freedom: polarization and direction of propagation, either to Alice (channel 2) or to Bob (channel 3). Their state vector is given by
\begin{gather}
\ket{\Psi^{(2,3)}} = \left(a_{V2,\,\omega}^\dag a_{H3,\,\omega}^\dag + a_{H2,\,\omega}^\dag a_{V3,\,\omega}^\dag\right) \ket{0}
\equiv
\label{SPDC-st-vect}
\ket{1_{H3,\,\omega},1_{V2,\,\omega}}+\ket{1_{H2,\,\omega},1_{V3,\,\omega}},
\end{gather}
where $a_{H2,\,\omega}^\dag$, $a_{H3,\,\omega}^\dag$,  $a_{V3,\,\omega}^\dag$ and $a_{V2,\,\omega}^\dag$ are the creation operators of photons with horizontal or vertical polarizations propagating in channel (2) or (3) and having the same frequency $\omega$. For shortening formulas, we drop in Eq. (\ref{SPDC-st-vect}) and below the normalizing factor $1/\sqrt{2}$ as not important for further consideration.

A light source of the message to be teleported to Bob via Alice serves for generation of an unknown quantum state with arbitrary polarization characterized by two complex constants $\alpha$ and $\beta$ such that $|\alpha|^2+|\beta|^2=1$. These constants can be used for encoding information for Bob. The propagation channel of the message photons is denoted as $(1)$, and the corresponding single-photon state vector is
\begin{equation}
\label{mess-st-vec}
\ket{\Psi^{(1)}}=(\alpha a_{H1,\,\omega'}^\dag + \beta a_{V1,\,\omega'}^\dag)\ket{0}\equiv\alpha\ket{1_{H1,\,\omega'}}+\beta\ket{1_{H1,\,\omega'}}.
\end{equation}
The frequency of message photons $\omega'$ is assumed to be pronouncedly different from both frequencies of SPDC photons and of the pump, $\omega^\prime\neq\omega,\,\,\omega_p$. For example, if the pump wavelength and the wavelength of SPDC photons are equal to $\lambda_p=0.5\,\mu{\rm m}$ and $\lambda_{\rm SPDC}=1\,\mu{\rm m}$, the wavelength of message photons can be taken equal to $\lambda_{\rm mes}=0.7\,\mu{\rm m}$. The state vector of all three indistinguishable photons participating in the QT process of Figure \ref{Fig.1} is given by :
\begin{equation}
\label{Three-phot st-vec}
\ket{\Psi^{(1,2,3)}}=(\alpha a_{H1,\,\omega^\prime}^\dagger + \beta a_{V1,\,\omega^\prime}^\dagger)(a_{V2,\,\omega}^\dagger a_{H3,\,\omega}^\dagger + a_{H2,\,\omega}^\dagger a_{V3,\,\omega}^\dagger) \ket{0}.
\end{equation}
Superscripts in state vectors $\ket{\Psi^{(...)}}$ here and below indicate involved propagation channels rather than numbers of photons.

Addition of the third photon of a different frequency to the SPDC pair extends the amount of photon's degrees of freedom up to three: polarization, propagation channels and frequencies. The wave function of such system completely symmetrized with respect to photon variables is inevitably very cumbersome and hardly convenient for further analysis. In contrast, the state vector of the system (\ref{Three-phot st-vec}) and its further transformations are very simple and informative.

The upper dichroic mirror shown in Figure \ref{Fig.1} is used for the SPDC photon sent to Alice to merge the channel (1) after which these photons propagate together with message photons. After this transformation the state vector (\ref{Three-phot st-vec}) takes the form
\begin{gather}
\nonumber
\ket{\Psi^{(1,3)}}={\Big(}\alpha a_{H3}^\dagger a_{H1,\,\omega'}^\dagger a_{V1,\,\omega}^\dagger +
\alpha a_{V3}^\dagger a_{H1,\,\omega'}^\dagger a_{H1,\,\omega}^\dagger +\\
\label{after DM1}
\beta a_{H3}^\dagger a_{V1,\,\omega'}^\dagger a_{V1,\,\omega}^\dagger+
\beta a_{V3}^\dagger a_{V1,\,\omega'}^\dagger a_{H1,\,\omega}^\dagger{\Big )} \ket{0}.
\end{gather}
Each of four terms on the right-hand side of this equation determines different mutually correlated distributions of photons to be received by Bob and Alice. Bob gets single photons with weighting coefficients $\alpha$ or $\beta$ and with accidentally distributed between these two cases horizontal or vertical polarizations. For transforming this accidental distribution of polarizations into that occurring in the message (\ref{mess-st-vec}), Bob has to make corrections for making all photons coming to him with the weighting coefficient $\alpha$ being polarized horizontally, and for all photons coming with the weighting coefficient $\beta$ being polarized vertically. However Bob himself cannot recognize photons in which he has to make corrections because he does not know in advance anything about the weighting coefficients $\alpha$ or $\beta$. Thus, Bob needs instructions to be received from Alice.

As for Alice, she also does not know anything about the weighting coefficients $\alpha$ and $\beta$. But she knows that her SPDC photons and those coming the same time to Bob obligatory have different polarizations. And she can compare polarization of every her SPDC photon with polarization of a  coming to her simultaneously message photon. This information can be used for formulating instructions for Bob on what he has to do with his counterpart SPDC photons before sending them to the detector.

Comparison of polarizations by Alice can be accomplished by the method used in the work \cite{SPK}. Alice can send  pairs of photons from the channel (1) after the upper ${\rm DM}$ into a group of four differently oriented nonlinear crystals transforming pairs into single photons with a summed frequency $\omega^{\prime\prime}=\omega'+\omega$. Such frequency summation process is an inverse version of  SPDC processes. Regimes of the frequency-summation processes in four crystals can be chosen corresponding to four terms in brackets on the right-hand side of Equation (\ref{after DM1}): the first crystal from the left corresponding to the first term in brackets of Equation (\ref{after DM1}), the second - to the second, etc. The resulting transformation rules are given by
\begin{equation}
\label{summation}
\begin{array}{c}
\mathcal{N}^{\underline{\rm o}}\,1:\,{\rm Type\, II},\quad H^{(e)}\omega' + V^{(o)}\omega\Rightarrow H^{(e)}\omega''  \\
 \mathcal{N}^{\underline{\rm o}}\,2:{\rm Type\, I},\quad H^{(o)}\omega' + H^{(o)}\omega\Rightarrow V^{(e)}\omega''\\
 \mathcal{N}^{\underline{\rm o}}\,3:{\rm Type\, I},\quad V^{(o)}\omega' + V^{(o)}\omega\Rightarrow H^{(e)}\omega'' \\
 \mathcal{N}^{\underline{\rm o}}\,4:{\rm Type\, II},\quad V^{(e)}\omega' + H^{(o)}\omega\Rightarrow V^{(e)}\omega''
\end{array}
\end{equation}
Here and in Fig. \ref{Fig.1} the numbers ${\rm I}$ and ${\rm II}$ refer to the phase matching types, and the superscripts $(o)$ and $(e)$ refer to the ordinary and extraordinary waves in crystals. The first and forth crystals in a scheme of Fig. \ref{Fig.1} can be taken identical except the fourth crystal has to be turned for $90^o$ around the $z-$axis. The same is true for the pair of the second and third crystals.

In terms of creation operators of photons with summed frequencies, the total three-photon state vector (\ref{Three-phot st-vec}), (\ref{after DM1}) turns into the biphoton state vector of the form
\begin{gather}
\ket{\Psi^{(1,3)}}=(\alpha a_{H3}^\dag a_{H1,\,\omega^{\prime\prime}}^\dag +
\alpha a_{V3}^\dagger a_{V1,\,\omega''}^\dag +
\label{biph-doubled}
\beta a_{H3}^\dagger a_{H1,\,\omega''}^\dag+
\beta a_{V3}^\dag a_{V1,\,\omega''}^\dag )\ket{0}.
\end{gather}
Features of this expression and its importance will be clarified a little bit later. But for comparison of polarizations by Alice this is not sufficient, because Alice has to be able to differentiate cases when each of four options (\ref{summation}) is realized separately from others. Once again, this task can be accomplished by the method similar to that used in the work \cite{SPK},  consisting in installation of dichroic mirrors after each of four crystals in the scheme of Figure \ref{Fig.1} and in registration of reflected photons with the summed frequency $\omega^{''}$ by separate detectors after each ${\rm DM}$. For the state vector (\ref{biph-doubled}) this means splitting for four parts corresponding to channels a, b, c and d:
\begin{gather}
\ket{\Psi^{(3,abcd)}}=(\alpha a_{H3}^\dag a_{Ha,\,\omega^{\prime\prime}}^\dag +
\alpha a_{V3}^\dagger a_{Vb,\,\omega''}^\dag +
\label{biph-doubled-2}
\beta a_{H3}^\dagger a_{Hc,\,\omega''}^\dag+
\beta a_{V3}^\dag a_{Vd,\,\omega''}^\dag )\ket{0}.
\end{gather}
Knowing which detectors clicks, Alice can say immediately which term in the sum of four terms in the state vector (\ref{after DM1}) is responsible for this process and, hence, which polarizations had in each given case message and SPDC photons which came to Alice. If the clicking detector is $\mathcal{N}^{\underline{\rm o}}$ 1 or $\mathcal{N}^{\underline{\rm o}}$ 4, from the first and fourth lines in the transformation rules (\ref{summation}) Alice finds immediately that her message and SPDC photons had different polarizations and, consequently, the SPDC photons of Bob had the same polarizations as in the message. Alice can send this good news to Bob by a phone call (or by any other way) and Bob can quietly send his SPDC photons to his detector. Oppositely, in the cases of clicking detectors $\mathcal{N}^{\underline{\rm o}}$ 2 or  $\mathcal{N}^{\underline{\rm o}}$ 3, from the middle two lines in the transformation rules (\ref{summation}) Alice finds that in these cases polarizations of message and her SPDC photons coincided. This means that polarizations of SPDC photons obtained by Bob were wrong and, hence, they needed corrections. As soon as Alice realizes this, she calls to Bob and instructs him to change polarizations of photons before sending them to the detector: $V\rightarrow H$ in the case of the clicking detector $\mathcal{N}^{\underline{\rm o}}\,2$ and  $H\rightarrow V$ if the clicking detector is $\mathcal{N}^{\underline{\rm o}}\,3$. This solves the task of accomplishing QT because after these corrections all Bob's photons will have the same polarizations and weighting coefficients $\alpha,\,\beta$ which occurred in the message.

If the corrections of polarizations made by Bob are taken into account in the $2nd$ and $3rd$ terms of the biphoton state vector (\ref{biph-doubled}), the latter takes the form
\begin{equation}
\label{factorized}
\ket{\Psi^{(1,3)}_{\rm corrected}} = (\alpha a_{H3,\,\omega}^\dag + \beta a_{V3,\,\omega}^\dag) (a_{H1,\,\omega''}^\dag + a_{V1,\,\,\omega''}^\dag) \ket{0}.
\end{equation}
As it's seen clearly, the parts of the state vector corresponding to channels (1) and (3) are factorized similarly to what has occurred in the original three-photon state vector (\ref{Three-phot st-vec}). Moreover, the part corresponding to the channel (3) is identical to the original message state vector (\ref{mess-st-vec}). All this indicates clearly existence of the QT effect.

Though looking very nice, technically the state vector of the form (\ref{factorized}) does not correspond to any stage of the experiment described above. Indeed, Bob can perform his corrections of polarizations only after getting instructions from Alice. In her turn, Alice can formulate these instructions after she sees all detector clicking and, hence, after all photons with the frequency $\omega''$ are absorbed at the detectors. As all these photons cease to exist after such measurements, the only remaining photons are those coming to Bob from the SPDC source with polarizations appropriately corrected whenever it was necessary. So, the  state vector of the remaining photons is just the part of the state vector (\ref{factorized}) corresponding to the channel 3
\begin{equation}
\label{final}
\ket{\Psi^{(3)}}_{final} = (\alpha a_{H3,\,\omega}^\dag + \beta a_{V3,\,\omega}^\dag) \ket{0},
\end{equation}
which is identical to the original message (\ref{mess-st-vec}).

The last note concerns the dashed lines used in Figure \ref{Fig.1} for the crystals $\mathcal{N}^{\underline{\rm o}}$ 1 and $\mathcal{N}^{\underline{\rm o}}$ 2. This is done to show that in principle one can avoid using these two crystals at all. Indeed, information which Alice can get  from clicking of the first or last detectors and which she can report to Bob is rather trivial. It's simply ``everything is OK, don't do anything with this photons, just send them to your detector". Alternatively,  Alice and Bob can agree in advance that the phone call takes some specified time $\Delta t$, and if the call does not come during this time, Bob understands that these photons already have appropriate polarizations and sends these photons to the detector. Thus, the minimal amount of frequency-summing crystals is two, and they have to be tuned in a way providing the transformation rules in two middle lines of Equation (\ref{summation}).

\section{A simple classical analog of QT}

\begin{figure}[h]
\centering\includegraphics[width=12cm]{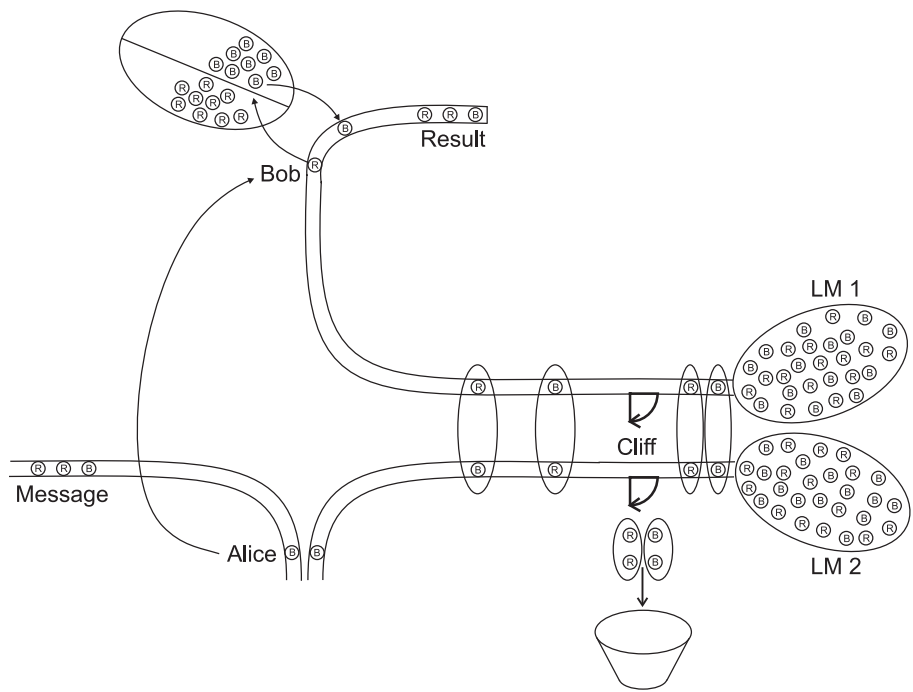}
\caption{{\protect\footnotesize {A scheme of the ``Classical QT".}}}\label{Fig.2}
\end{figure}

The described above explanation of QT, as well as most of other works and interpretations are based on quantum-electrodynamical or quantum optical-approaches. Evidently, some elements of these approaches are essentially quantum. Some of them concern the mentioned indistinguishability of quantum particles, states of which can be used for QT, and related to this obligatory symmetry features of the corresponding wave functions. The formalism of photon creation operators described above also is based on the use of concepts of quantum electrodynamics, and in this sense seems essentially quantum. But all this does not forbid asking question whether QT is an essentially quantum phenomenon or it's more classical than usually thought? In particular, these questions were raised by D.N. Klyshko \cite{Klyshko1}, \cite{Klyshko2}, who suggested some classical analogs of QT based on the use of classical light beams and classical optical devices. Below we describe a very simple absolutely classical scheme not using either photons or light beams but looking strikingly similar to the above described scheme of QT. The scheme is shown in  Figure \ref{Fig.2}.

The scheme includes two lottery machines ($\rm LM$), each filled with equal amounts of red (R) and blue (B) balls. From time to time windows of LMs open simultaneously for a short time sufficient for one ball to escape from each LM. Colors of escaping balls are accidental. After escaping from LM, the balls roll in pairs, each along its own chute. In Figure \ref{Fig.2} pairs of balls are surrounded by oval curves. There are four possible combinations of colors in pairs of balls: RB, BR, RR, and BB. The pairs of balls with coinciding colors differ significantly the arising sequence of balls from the quantum-electrodynamical state of photons with different polarizations (\ref{SPDC-st-vect}). But this difference can be easily removed if Alice and Bob invite a third participant, Cliff. The latter has to watch arriving pairs of balls and to push a button for opening windows in both chutes when in coming pairs of balls their colors are identical. Such balls fall down through the opened windows and in this way they are removed from the sequence of balls rolling further along their chutes directed finally to Alice and to Bob, as  shown in Figure \ref{Fig.2}. This sequence of balls consists now of accidentally changing themselves pairs RB and BR, and this is a complete classical analog of the entangled quantum-electrodynamical state of Equation  (\ref{SPDC-st-vect}). Both Alice and Bob receive one of two balls of each pair, and they know exactly that colors of balls coming to each of them are different: if Alice gets a red ball she knows that the ball received by Bob is blue, and vice versa.

A message, that has to be transferred to Bob via Alice can be encoded in different sequences of balls of identical or different colors. E.g., in the manifold of three balls there are 8 different combinations: RRR, RRB, RBR, BRR, RBB, BRB, BBR, BBB, which can represent 8 elements of the encoding alphabet. Specific example of a message shown in Figure \ref{Fig.2} is RRB. Alice does not know either the meaning of these symbols or of the message to be sent to Bob. But she can compare colors of balls coming to her from the message source and from the LM 2. If Alice sees that these balls have different colors, she conclude immediately that the ball received in the same time by Bob has the same color as that of the message element seen by Alice. She can call Bob by phone to inform him about this conclusion, and then Bob can collect the ball received by him from LM 1 in a special box as an element of the message he plans to read. However, if Alice sees that colors of her two balls are identical, she realizes that Bob receives in the same time a ball of a wrong color. Alice informs Bob and he replaces the wrong-color ball received from LM 1 by a ball of a different color after which he saves this replacement ball in the box for copying the message. After repeating these procedures several times, Bob gets finally in his message box the exact copy of the several-ball encoded message produced in the message source. To read this message, of course, Bob has to know the alphabet used for encoding it, which is assumed to be arranged in advance.

\section{Conclusion}

Two main conclusions correspond to two sections of the paper. First, the description in terms of the photon creation operators is absolutely clear and very simple. It consists of a few simple formulas from (\ref{SPDC-st-vect}) to (\ref{factorized}), and the last Equation (\ref{factorized}) shows that  after summation of frequencies the final state vector takes the factorized form of a product of creation operators of photons in the Bob's (3) and Alice's (1) channels. And the part of creations operators in the channel of Bob is identical to that occurring in the original message state vector (\ref{mess-st-vec}). And the second conclusion following directly from the section 3 concerns the question about classicality vs quantumness of QT. The described example of an absolutely classical and easily doable experiment shows that in this scheme all operations with classical objects (balls of different colors) practically repeat operations with photons of different polarizations, and the final result is the same as in QT: reproduction of the message by Bob with the help of Alice. In our opinion this example shows clearly that the essence of QT is classical and its quantumness is related only with the used objects. As photons are quanta of light, in this sense they are  quantum objects which justifies the name quantum teleportation. But the processes itself and involved transformations in QT are practically the same as in the classical analog and hardly can pretend for being exclusively quantum ones. It may be reasonable saying that QT is a classical phenomenon most often using quantum objects for its realization.

\section*{Acknowledgement}
The work is supported by the Russian Science Foundation, grant 14-02-01338-$\Pi$

\bibliographystyle{unsrtnat}

\bibliography{text}

\begin{thebibliography}{8}

\bibitem{Bennett}
C.~H. Bennett, G.~Brassard, C.~Cr${\rm\acute{e}}$peau, R.~Jozsa, A.~Peres, and
  W.~K. Wootters.
\newblock \emph{Phys.Rev.Lett}, 70:1895--1899, 1993.

\bibitem{Zeilinger}
D.~Boumeester, J.-W. Pan, K.~Mattle, M.~Eibl, H.~Weinfurter, and A.~Zeilinger.
\newblock \emph{Nature}, 390:575--579, 1997.

\bibitem{DeMartini}
D.~Boschi, S.~Branca, F.~De Martini, L.~Hardy, and S.~Popescu.
\newblock \emph{Phys. Rev. Lett.}, 80:1121--1125, 1998.

\bibitem{SPK}
Y.-H. Kim, S.~P. Kulik, and Y.~Shih.
\newblock \emph{Phys. Rev. Lett.}, 86:1370, 2001.

\bibitem{Schwe}
S.~S. Schweber.
\newblock \emph{An Introduction to relativistic quantum field theory}.
\newblock Harper \& Row Publishers, New York, USA, 1961.

\bibitem{Kwiat}
P.~G. Kwiat, K.~Mattle, H.~Weinfurter, A.~Zeilinger, A.~V. Sergienko, and
  Y.~Shih.
\newblock \emph{Phys. Rev. Lett.}, 75:4337, 1995.

\bibitem{Klyshko1}
D.~N. Klyshko.
\newblock \emph{Journal of Experimental and Theoretical Physics}, 87:639, 1998.

\bibitem{Klyshko2}
D.~N. Klyshko.
\newblock \emph{Phys. Lett. A}, 247:261, 1998.

\end{thebibliography}

\end{document}